\pdfoutput=1
\documentclass[12pt,a4paper]{article}
\usepackage{ifthen} % for conditional statements
\newboolean{pdflatex}
\setboolean{pdflatex}{true} % False for eps figures 

\newboolean{articletitles}
\setboolean{articletitles}{true} % False removes titles in references

\newboolean{uprightparticles}
\setboolean{uprightparticles}{false} %True for upright particle symbols

\def\paperauthors{LHCb collaboration} % Leave as is for PAPER, CONF and FIGURE
\def\paperasciititle{Associated production of prompt Jpsi and Upsilon mesons in pp collisions at 13 TeV} % Set ASCII title here !! MAKE sure it's only ASCII characters !! 
\def\papertitle{Associated production of prompt \jpsi and \Upsilonres mesons in \proton\proton collisions at $\sqs=13\tev$} % Latex formatted title
\def\paperkeywords{{High Energy Physics}, {LHCb}} % Comma separated list
\def\papercopyright{\the\year\ CERN for the benefit of the LHCb collaboration} % new since 9/Apr/2018
\def\paperlicence{CC BY 4.0 licence}
\def\paperlicenceurl{https://creativecommons.org/licenses/by/4.0/}

\usepackage[top=1in, bottom=1.25in, left=1in, right=1in]{geometry}

\usepackage{microtype}
\usepackage{lineno}  % for line numbering during review
\usepackage{xspace} % To avoid problems with missing or double spaces after
\usepackage{caption} %these three command get the figure and table captions automatically small

\usepackage{graphicx}  % to include figures (can also use other packages)
\usepackage{color}
\usepackage{colortbl}
\graphicspath{{./figs/}} % Make Latex search fig subdir for figures
\usepackage{amsmath} % Adds a large collection of math symbols
\usepackage{amssymb}
\usepackage{amsfonts}
\usepackage{upgreek} % Adds in support for greek letters in roman typeset

\newcommand*\patchAmsMathEnvironmentForLineno[1]{%
\expandafter\let\csname old#1\expandafter\endcsname\csname #1\endcsname
\expandafter\let\csname oldend#1\expandafter\endcsname\csname
end#1\endcsname
 \renewenvironment{#1}%
   {\linenomath\csname old#1\endcsname}%
   {\csname oldend#1\endcsname\endlinenomath}%
}
\newcommand*\patchBothAmsMathEnvironmentsForLineno[1]{%
  \patchAmsMathEnvironmentForLineno{#1}%
  \patchAmsMathEnvironmentForLineno{#1*}%
}
\AtBeginDocument{%
\patchBothAmsMathEnvironmentsForLineno{equation}%
\patchBothAmsMathEnvironmentsForLineno{align}%
\patchBothAmsMathEnvironmentsForLineno{flalign}%
\patchBothAmsMathEnvironmentsForLineno{alignat}%
\patchBothAmsMathEnvironmentsForLineno{gather}%
\patchBothAmsMathEnvironmentsForLineno{multline}%
\patchBothAmsMathEnvironmentsForLineno{eqnarray}%
}

\usepackage{hyperxmp}

\usepackage[pdftex,
            pdfauthor={\paperauthors},
            pdftitle={\paperasciititle},
            pdfkeywords={\paperkeywords},
            pdfcopyright={Copyright (C) \papercopyright},
            pdflicenseurl={\paperlicenceurl}]{hyperref}
\usepackage[colorinlistoftodos,textsize=scriptsize]{todonotes}

\usepackage[bottom,flushmargin,hang,multiple]{footmisc}

\usepackage[all]{hypcap} % Internal hyperlinks to floats.

\usepackage{xspace} 
\usepackage{upgreek}

\def\lhcb   {\mbox{LHCb}\xspace}

\def\dzero  {\mbox{D0}\xspace}

\def\MagUp {\mbox{\em Mag\kern -0.05em Up}\xspace}

\ifthenelse{\boolean{uprightparticles}}%
{

 \def\Pmu         {\ensuremath{\upmu}\xspace}

 \def\Ppi         {\ensuremath{\uppi}\xspace}

 \def\Ppsi        {\ensuremath{\uppsi}\xspace}

 \def\PDelta      {\ensuremath{\Delta}\xspace}                 
 \def\PXi         {\ensuremath{\Xi}\xspace}                 
 \def\PLambda     {\ensuremath{\Lambda}\xspace}                 
 \def\PSigma      {\ensuremath{\Sigma}\xspace}                 
 \def\POmega      {\ensuremath{\Omega}\xspace}                 
 \def\PUpsilon    {\ensuremath{\Upsilon}\xspace}
 \let\oldPi\Pi
 \def\PPi         {\ensuremath{\oldPi}\xspace}

 \def\PB      {\ensuremath{\mathrm{B}}\xspace}                 
                  
 \def\PD      {\ensuremath{\mathrm{D}}\xspace}

 \def\PJ      {\ensuremath{\mathrm{J}}\xspace}                 
 \def\PK      {\ensuremath{\mathrm{K}}\xspace}

 \def\Pb      {\ensuremath{\mathrm{b}}\xspace}                 
 \def\Pc      {\ensuremath{\mathrm{c}}\xspace}

 \def\Pi      {\ensuremath{\mathrm{i}}\xspace}

 \def\Pp      {\ensuremath{\mathrm{p}}\xspace}

 \def\Ps      {\ensuremath{\mathrm{s}}\xspace}

 \def\thebaroffset{0.0em}
}
{

 \def\Pmu         {\ensuremath{\mu}\xspace}

 \def\Ppi         {\ensuremath{\pi}\xspace}

 \def\Ppsi        {\ensuremath{\psi}\xspace}                 
                  
 \mathchardef\PDelta="7101
 \mathchardef\PXi="7104
 \mathchardef\PLambda="7103
 \mathchardef\PSigma="7106
 \mathchardef\POmega="710A
 \mathchardef\PUpsilon="7107
 \mathchardef\PPi="7105
                  
 \def\PB      {\ensuremath{B}\xspace}                 
                  
 \def\PD      {\ensuremath{D}\xspace}

 \def\PJ      {\ensuremath{J}\xspace}                 
 \def\PK      {\ensuremath{K}\xspace}

 \def\Pb      {\ensuremath{b}\xspace}                 
 \def\Pc      {\ensuremath{c}\xspace}

 \def\Pi      {\ensuremath{i}\xspace}

 \def\Pp      {\ensuremath{p}\xspace}

 \def\Ps      {\ensuremath{s}\xspace}

 \def\thebaroffset{0.18em}
}
\newcommand{\offsetoverline}[2][\thebaroffset]{\kern #1\overline{\kern -#1 #2}}%

\makeatletter
\ifcase \@ptsize \relax% 10pt
  \newcommand{\miniscule}{\@setfontsize\miniscule{4}{5}}% \tiny: 5/6
\or% 11pt
  \newcommand{\miniscule}{\@setfontsize\miniscule{5}{6}}% \tiny: 6/7
\or% 12pt
  \newcommand{\miniscule}{\@setfontsize\miniscule{5}{6}}% \tiny: 6/7
\fi
\makeatother

\DeclareRobustCommand{\optbar}[1]{\shortstack{{\miniscule (\rule[.5ex]{1.25em}{.18mm})}
  \\ [-.7ex] $#1$}}

\def\mup        {{\ensuremath{\Pmu^+}}\xspace}
\def\mun        {{\ensuremath{\Pmu^-}}\xspace} % muon negative (\mum is taken)

\def\mumu       {{\ensuremath{\Pmu^+\Pmu^-}}\xspace}

\def\squark    {{\ensuremath{\Ps}}\xspace}

\def\cquark    {{\ensuremath{\Pc}}\xspace}

\def\bquark    {{\ensuremath{\Pb}}\xspace}

\def\pion   {{\ensuremath{\Ppi}}\xspace}

\def\KorKbar {\kern \thebaroffset\optbar{\kern -\thebaroffset \PK}{}\xspace}

\def\D       {{\ensuremath{\PD}}\xspace}

\def\DorDbar {\kern \thebaroffset\optbar{\kern -\thebaroffset \PD}\xspace}

\def\Dp      {{\ensuremath{\D^+}}\xspace}
\def\Dm      {{\ensuremath{\D^-}}\xspace}

\def\DpDm    {\ensuremath{\Dp {\kern -0.16em \Dm}}\xspace}

\def\B       {{\ensuremath{\PB}}\xspace}

\def\BorBbar {\kern \thebaroffset\optbar{\kern -\thebaroffset \PB}\xspace}

\def\Bd      {{\ensuremath{\B^0}}\xspace}

\def\BdorBdbar {\kern \thebaroffset\optbar{\kern -\thebaroffset \Bd}\xspace}

\def\Bs      {{\ensuremath{\B^0_\squark}}\xspace}

\def\BsorBsbar {\kern \thebaroffset\optbar{\kern -\thebaroffset \Bs}\xspace}

\def\jpsi     {{\ensuremath{{\PJ\mskip -3mu/\mskip -2mu\Ppsi}}}\xspace}

\def\Upsilonres  {{\ensuremath{\PUpsilon}}\xspace}
\def\Y#1S{\ensuremath{\PUpsilon{(#1S)}}\xspace}
\def\OneS  {{\Y1S}\xspace}
\def\TwoS  {{\Y2S}\xspace}
\def\ThreeS{{\Y3S}\xspace}

\def\proton      {{\ensuremath{\Pp}}\xspace}

\def\LorLbar     {\kern \thebaroffset\optbar{\kern -\thebaroffset \PLambda}\xspace}

\def\BF         {{\ensuremath{\mathcal{B}}}\xspace}

\def\to                 {\ensuremath{\rightarrow}\xspace}

\def\AT#1     {\ensuremath{A_{\mathrm{T}}^{#1}}\xspace}           % 2

\def\C#1      {\ensuremath{\mathcal{C}_{#1}}\xspace}                       % 9
\def\Cp#1     {\ensuremath{\mathcal{C}_{#1}^{'}}\xspace}                    % 7
\def\Ceff#1   {\ensuremath{\mathcal{C}_{#1}^{\mathrm{(eff)}}}\xspace}        % 9  
\def\Cpeff#1  {\ensuremath{\mathcal{C}_{#1}^{'\mathrm{(eff)}}}\xspace}       % 7
\def\Ope#1    {\ensuremath{\mathcal{O}_{#1}}\xspace}                       % 2
\def\Opep#1   {\ensuremath{\mathcal{O}_{#1}^{'}}\xspace}                    % 7

\newcommand{\nospaceunit}[1]{\ensuremath{\text{#1}}}       
\newcommand{\aunit}[1]{\ensuremath{\text{\,#1}}}       
\newcommand{\tev}{\aunit{Te\kern -0.1em V}\xspace}
\newcommand{\gev}{\aunit{Ge\kern -0.1em V}\xspace}
\newcommand{\mev}{\aunit{Me\kern -0.1em V}\xspace}
\newcommand{\kev}{\aunit{ke\kern -0.1em V}\xspace}
\newcommand{\ev}{\aunit{e\kern -0.1em V}\xspace}
 
\newcommand{\mevc}{\ensuremath{\aunit{Me\kern -0.1em V\!/}c}\xspace}
\newcommand{\gevc}{\ensuremath{\aunit{Ge\kern -0.1em V\!/}c}\xspace}
\newcommand{\mevcc}{\ensuremath{\aunit{Me\kern -0.1em V\!/}c^2}\xspace}
\newcommand{\gevcc}{\ensuremath{\aunit{Ge\kern -0.1em V\!/}c^2}\xspace}
\newcommand{\gevgevcc}{\ensuremath{\gev^2\!/c^2}\xspace} % for \pt^2 in CEP
\def\mum  {\ensuremath{\,\upmu\nospaceunit{m}}\xspace}

\def\mbarn{\aunit{mb}\xspace}

\def\pb {\aunit{pb}\xspace}

\def\fb   {\ensuremath{\aunit{fb}}\xspace}
\def\invfb   {\ensuremath{\fb^{-1}}\xspace}

\newcommand{\stat}{\aunit{(stat)}\xspace}
\newcommand{\syst}{\aunit{(syst)}\xspace}

\def\gsim{{~\raise.15em\hbox{$>$}\kern-.85em
          \lower.35em\hbox{$\sim$}~}\xspace}
\def\lsim{{~\raise.15em\hbox{$<$}\kern-.85em
          \lower.35em\hbox{$\sim$}~}\xspace}

\def\sPlot{\mbox{\em sPlot}\xspace}

\def\sqs   {\ensuremath{\protect\sqrt{s}}\xspace}

\def\pt         {\ensuremath{p_{\mathrm{T}}}\xspace}

\def\ptot       {\ensuremath{p}\xspace}

\def\geant      {\mbox{\textsc{Geant4}}\xspace}

\def\tell1  {TELL1\xspace}
\def\ukl1   {UKL1\xspace}

\newcommand{\eg}{\mbox{\itshape e.g.}\xspace}
\newcommand{\ie}{\mbox{\itshape i.e.}\xspace}

\newcommand{\lhcborcid}[1]{\href{https://orcid.org/#1}{\hspace*{0.1em}\raisebox{-0.45ex}{\includegraphics[width=1em]{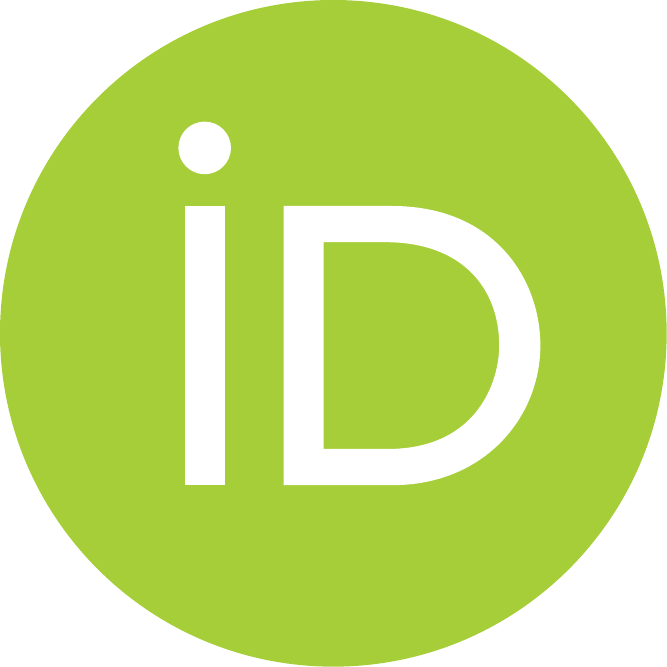}}}}

\usepackage{cite} % Allows for ranges in citations
\usepackage{mciteplus}

\newcommand{\XsecOneS}[0]{133}
\newcommand{\SigmaStatXsecOneS}[0]{22}
\newcommand{\SigmaSystXsecOneS}[0]{7}
\newcommand{\SigmaExtXsecOneS}[0]{3}

\newcommand{\XsecTwoS}[0]{76}
\newcommand{\SigmaStatXsecTwoS}[0]{21}
\newcommand{\SigmaSystXsecTwoS}[0]{4}
\newcommand{\SigmaExtXsecTwoS}[0]{7}

\newcommand{\SigmaOne}[0]{$7.9\,\sigma$\xspace}
\newcommand{\SigmaTwo}[0]{$4.9\,\sigma$\xspace}

\newcommand{\NcorOneSInclusive}[0]{840}
\newcommand{\DeltaNcorOneSInclusive}[0]{140}

\newcommand{\NcorTwoSInclusive}[0]{370}
\newcommand{\DeltaNcorTwoSInclusive}[0]{100}

\newcommand{\SigOne}[0]{$\jpsi$--$\OneS$\xspace}
\newcommand{\SigTwo}[0]{$\jpsi$--$\TwoS$\xspace}

\def\JpsiUpsilon     {\ensuremath{ \jpsi\text{--}\PUpsilon}\xspace}

\begin{document}

%%%%%%%%%%%%%%%%%%%%%%%%%
%%%%% Title     %%%%%%%%%
%%%%%%%%%%%%%%%%%%%%%%%%%
\renewcommand{\thefootnote}{\fnsymbol{footnote}}
\setcounter{footnote}{1}

% %%%%%%% CHOOSE TITLE PAGE--------
%\onecolumn
%\input{title-LHCb-INT}
%\input{title-LHCb-ANA}
%\input{title-LHCb-CONF}
%\input{title-LHCb-FIGURE}
% ===============================================================================
% Purpose: LHCb-PAPER journal paper title page template
% Author: 
% Created on: 2010-09-25
% ===============================================================================

%%%%%%%%%%%%%%%%%%%%%%%%%
%%%%%  TITLE PAGE  %%%%%%
%%%%%%%%%%%%%%%%%%%%%%%%%
\begin{titlepage}
\pagenumbering{roman}

% Header ---------------------------------------------------
\vspace*{-1.5cm}
\centerline{\large EUROPEAN ORGANIZATION FOR NUCLEAR RESEARCH (CERN)}
\vspace*{1.5cm}
\noindent
\begin{tabular*}{\linewidth}{lc@{\extracolsep{\fill}}r@{\extracolsep{0pt}}}
\ifthenelse{\boolean{pdflatex}}% Logo format choice
{\vspace*{-1.5cm}\mbox{\!\!\!\includegraphics[width=.14\textwidth]{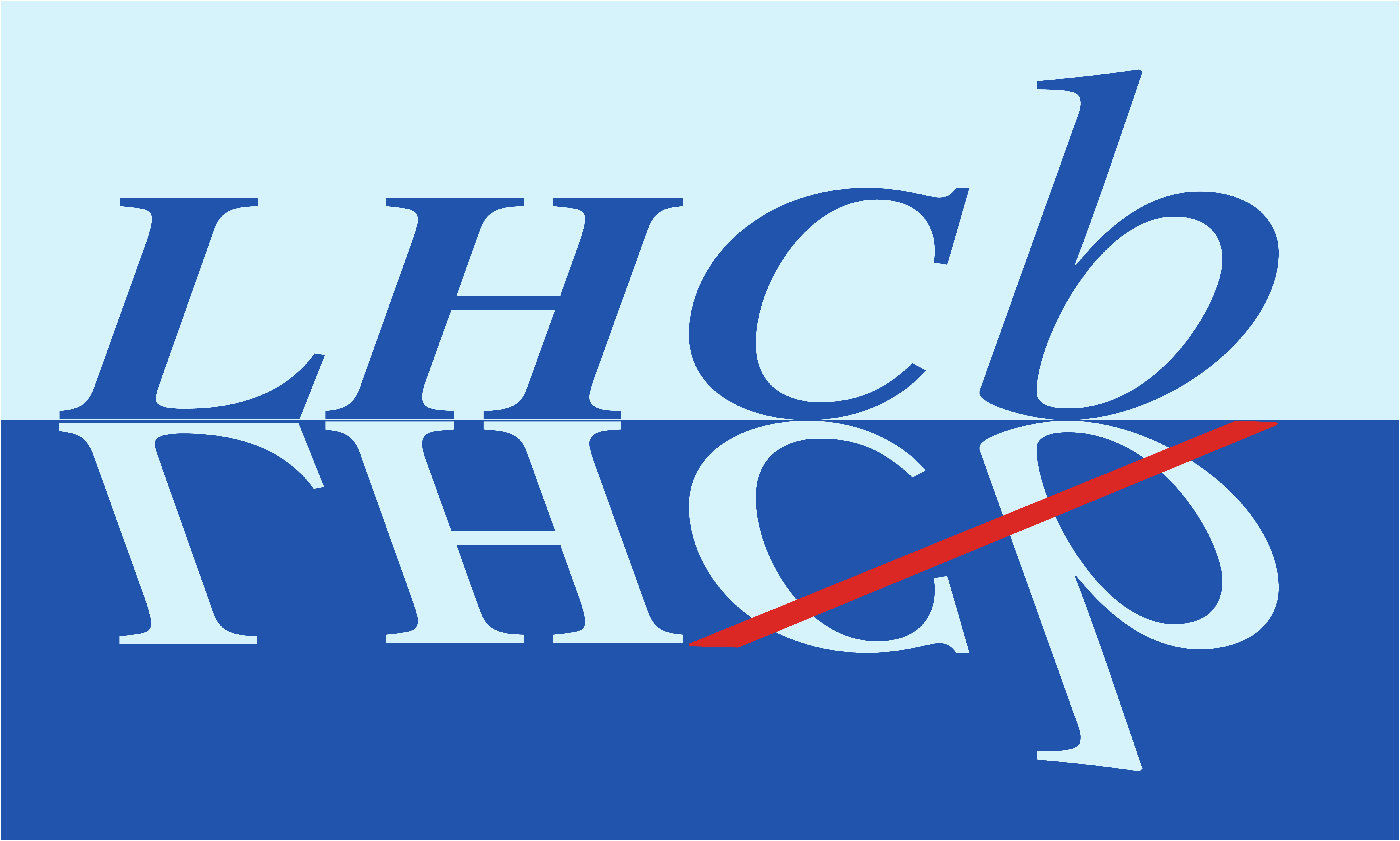}} & &}%
{\vspace*{-1.2cm}\mbox{\!\!\!\includegraphics[width=.12\textwidth]{figs/lhcb-logo.eps}} & &}%
\\
 & & CERN-EP-2023-078 \\  % ID 
 & & LHCb-PAPER-2022-047 \\  % ID 
 & & August 29, 2023 \\ % Date - Can also hardwire e.g.: 23 March 2010
 & & \\
% not in paper \hline
\end{tabular*}

\vspace*{2cm}

% Title --------------------------------------------------
{\normalfont\bfseries\boldmath\huge
\begin{center}
% DO NOT EDIT HERE. Instead edit macro in main.tex to keep metadata correct
  \papertitle 
\end{center}
}

\vspace*{2.0cm}

% Authors -------------------------------------------------
\begin{center}
%In the footnote, replace 'paper' by 'Letter' in case of submission to PRL or PLB 
% Edit macro in main.tex to keep metadata correct
\paperauthors\footnote{Authors are listed at the end of this paper.}
\end{center}

\vspace{\fill}

% Abstract -----------------------------------------------
\begin{abstract}
   \noindent
The associated production of prompt \jpsi and $\Upsilonres$ mesons in $\proton\proton$ collisions at a centre-of-mass energy of  $\sqs=13\tev$ is studied using LHCb data, corresponding to an integrated luminosity of $4\invfb$.
The measurement is performed for \jpsi (\Upsilonres) mesons with a transverse momentum $\pt<10\text{ }(30)\gevc$ in the rapidity range $2.0<y<4.5$.
In this kinematic range, the cross-section of the associated production of prompt \jpsi and $\OneS$ mesons is measured to be $\XsecOneS \pm \SigmaStatXsecOneS \pm \SigmaSystXsecOneS  \pm \SigmaExtXsecOneS \pb$, with a significance of \SigmaOne, and that of prompt \jpsi and $\TwoS$ mesons to be $\XsecTwoS \pm \SigmaStatXsecTwoS  \pm\SigmaSystXsecTwoS  \pm\SigmaExtXsecTwoS \pb$, with a significance of \SigmaTwo. The first uncertainty is statistical, the second systematic, and the third due to uncertainties on the used branching fractions.
This is the first observation of the associated production of \jpsi and \OneS in proton-proton collisions.
Differential cross-sections are measured as functions of variables that are sensitive to kinematic correlations between the \jpsi and \OneS mesons.
The effective cross-sections of the associated production of prompt \jpsi and \Upsilonres mesons 
are obtained and found to be compatible with measurements using other particle productions.

\end{abstract}

\vspace*{2.0cm}

\begin{center}
 Published in JHEP 08 (2023) 093
\end{center}

\vspace{\fill}

{\footnotesize 
% Edit macro in main.tex to keep metadata correct
\centerline{\copyright~\papercopyright. \href{\paperlicenceurl}{\paperlicence}.}}
\vspace*{2mm}

\end{titlepage}

%%%%%%%%%%%%%%%%%%%%%%%%%%%%%%%%
%%%%%  EOD OF TITLE PAGE  %%%%%%
%%%%%%%%%%%%%%%%%%%%%%%%%%%%%%%%

%  empty page follows the title page ----
\newpage
\setcounter{page}{2}
\mbox{~}
%\newpage
%
%% Author List ----------------------------
%%  You need to get a new author list!
%\input{Authorship_LHCb-PAPER-2022-047.tex}
%
%The author list for journal publications is provided by the Membership Committee shortly after 'approval to go to paper' has been given.
%%It will be made available on the page
%%\verb!http://www.physik.uzh.ch/~strauman/forMemCo/LHCb-PAPER-XXXX-XXX/! .
%It will be sent to you by email shortly after a paper number has beens assigned.
%The author list should be included already at first circulation, 
%to allow new members of the collaboration to verify whether they have been included correctly.
%Occasionally a misspelled name is corrected or associated institutions become full members.
%In that case, a new author list will be sent to you.
%In case line numbering doesn't work well after including the authorlist, try moving the \verb!\bigskip! after the last author to a separate line.
%
%
%The authorship for Conference Reports should be ``The LHCb
%  collaboration'', with a footnote giving the name(s) of the contact
%  author(s), but without the full list of collaboration names.

%\twocolumn
% %%%%%%%%%%%%% ---------

\renewcommand{\thefootnote}{\arabic{footnote}}
\setcounter{footnote}{0}

%%%%%%%%%%%%%%%%%%%%%%%%%%%%%%%%
%%%%%  Table of Content   %%%%%%
%%%%%%%%%%%%%%%%%%%%%%%%%%%%%%%%
%%%% Uncomment if desired
%\tableofcontents
\cleardoublepage

%%%%%%%%%%%%%%%%%%%%%%%%%
%%%%% Main text %%%%%%%%%
%%%%%%%%%%%%%%%%%%%%%%%%%

\pagestyle{plain} % restore page numbers for the main text
\setcounter{page}{1}
\pagenumbering{arabic}

%% Uncomment during review phase. 
%% Comment before a final submission.
%\linenumbers

%% This is the main body
%% It is useful to have a single file so comemnts are not missed in overleaf.
\section{Introduction}
\label{sec:intro}
Hadroproduction of heavy quarkonia has been extensively studied to probe quantum chromodynamics (QCD)~\cite{Chapon:2020heu}. Various theoretical treatments have been proposed that model the production of quark-antiquark pairs $Q\overline{Q}$ and their subsequent hadronisation into quarkonia.  Promising approaches 
include the colour-singlet model (CSM)~\cite{Chang:1979nn,Berger:1980ni,Baier:1981uk} and the non-relativistic QCD (NRQCD) framework~\cite{Bodwin:1994jh}. The CSM considers only the colour-singlet intermediate $Q\overline{Q}$ state with the same $J^{PC}$ quantum numbers as the final-state quarkonium, while the NRQCD framework considers both colour-singlet and colour-octet $Q\overline{Q}$ states. 
The NRQCD framework assumes process-independent long-distance matrix elements (LDMEs) to describe the transition probabilities from the intermediate $Q\overline{Q}$ states to the final-state quarkonium. 
Quarkonium hadroproduction cross-sections as a function of the transverse momentum~($\pt$) are successfully described by NRQCD calculations~\cite{Bodwin:2014gia,Braaten:1994xb,Cho:1995vh,Ma:2010yw,Gong:2012ug, Butenschoen:2010rq}. However, there are still challenges in providing a coherent description of all heavy quarkonium hadroproduction data, \eg there is no unified description of the cross-sections and polarisation of the vector quarkonium states; and the inconsistency between the LDMEs of the $\eta_c$ and $\jpsi$ states---which are related to the heavy-quark spin symmetry---remains to be understood~\cite{Shao:2014yta,Brambilla:2010cs, Han:2014jya,Chao:2012iv,Zhang:2014ybe,Butenschoen:2014dra,Braaten:1999qk, LHCB-PAPER-2016-064}.

\begin{figure}[!b]
    \centering
    \includegraphics[width = 0.35\textwidth]{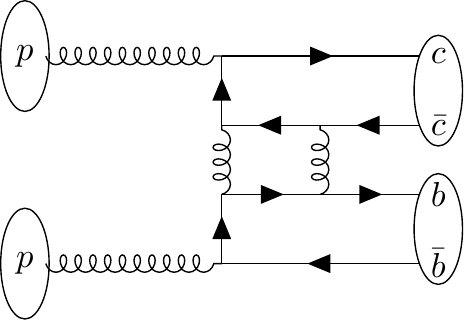}
    \qquad
    \includegraphics[width = 0.35\textwidth]{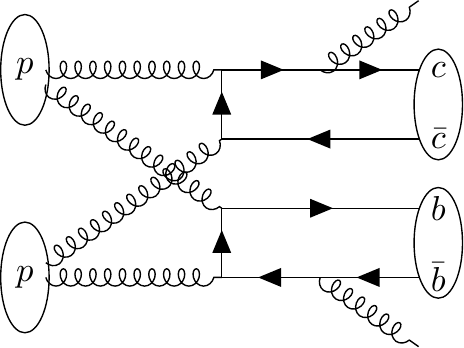}
    \caption{Representative Feynman diagrams of the (left) SPS and the (right) DPS process with colour-singlet LDMEs.}
    \label{fig:Feynman}
\end{figure}

In hadron collisions, a pair of heavy quarkonia can be produced through either single-parton scattering (SPS), or double-parton scattering (DPS), depending on the number of parton-parton scatterings in the process. 
A representative Feynman diagram of each process is presented in Fig.~\ref{fig:Feynman}. 
In the SPS process, two quarkonia are produced in the same parton-parton interaction, leading to  kinematic correlations between them.   In contrast, two parton-parton interactions occur simultaneously and independently in the DPS process,  producing two  quarkonia with weak kinematic correlations.
While the SPS process provides new insights into the heavy quarkonium production mechanism according to the left Feynman diagram in Fig.~\ref{fig:Feynman}~\cite{Shao:2016wor,Sun:2014gca,Li:2013csa,He:2018hwb},
the DPS process probes the parton distribution function inside the colliding hadrons and provides valuable information on the hadronic wave functions describing correlations among partons~\cite{Chapon:2020heu}. 
Taking the $\jpsi$ and $\PUpsilon$ associated production (referred to as \JpsiUpsilon production throughout) as an example\footnote{Throughout the paper, $\Upsilonres$ represents $\OneS$, $\TwoS$, or $\ThreeS$.}, the DPS cross-section $\sigma_{\text{DPS}}(\JpsiUpsilon)$ can be estimated from the $\jpsi$ and $\PUpsilon$ cross-sections, $\sigma(\jpsi)$ and $\sigma(\Upsilonres)$,  as
\begin{equation}
\label{equ:effective}
    \sigma_{\text{DPS}}(\JpsiUpsilon)= \frac{\sigma(\jpsi) \times \sigma(\Upsilonres)}{\sigma_{\mathrm{eff}}},
\end{equation}
where 
$\sigma_{\mathrm{eff}}$ is the effective cross-section parameter encoding the transverse overlapping region between partons in the protons~\cite{PhysRevD.57.503,PhysRevD.60.054023,PhysRevD.63.057901}. 
The effective cross-section is expected to be universal based on the assumptions that the two sets of colliding partons are uncorrelated and the longitudinal and transverse components of the parton distribution function factorise.
Measurements in different processes for the effective cross-section are mostly within the range of 5 to $35\mbarn$~\cite{Belyaev:2017sws,Lansberg:2016rcx}.
Further study is needed to test its universality~\cite{Lansberg:2019adr}.

A range of associated production processes of heavy quarkonia has been studied experimentally~\cite{Lansberg:2014swa,Lansberg:2016rcx,Lansberg:2017chq,Lansberg:2016muq,D0:2014vql,ATLAS:2016ydt,Shao:2016wor,ATLAS:2013aph,CMS:2013huw,LHCB-PAPER-2012-003,LHCB-PAPER-2015-046,CDF:1993sbj,CDF:1997yfa,D0:2009apj,Shao:2020kgj}. 
Heavy-quarkonium-pair production was first observed by the NA3 collaboration in the study of \jpsi-pair production in $\pion^-$-induced interactions~\cite{NA3:1982qlq}.
Subsequently, \jpsi-pair production was measured by the LHCb experiment~\cite{LHCB-PAPER-2011-013} and the CMS experiment~\cite{CMS:2014cmt} at $\sqs = 7 \tev$, by the \dzero collaboration  at $\sqs = 1.96 \tev$ in $p\bar{p}$ collisions~\cite{D0:2014vql}, by the ATLAS experiment at $\sqs = 8 \tev$~\cite{ATLAS:2016ydt}, and by the LHCb experiment at $\sqs = 13 \tev$~\cite{LHCB-PAPER-2016-057}.
In the di-$\jpsi$ invariant-mass spectrum, a structure of fully charmed tetraquark states was observed by LHCb~\cite{LHCb-PAPER-2020-011}, and recently confirmed by CMS~\cite{CMS:2023owd} and ATLAS~\cite{ATLAS:2023bft}. 
Production of \Upsilonres pairs was observed by the CMS experiment in $pp$ collisions at $\sqs = 8 \tev$~\cite{CMS:2016liw}.
For \JpsiUpsilon production, 
experimental evidence was found by the \dzero collaboration~\cite{D0:2015dyx}, but to date, no significant signals have been reported.

In this paper, \JpsiUpsilon production in $pp$ collisions at a centre-of-mass energy of $\sqs=13\tev$ is studied for \jpsi (\Upsilonres) mesons with a transverse momentum $\pt<10\text{ }(30)\gevc$ in the rapidity range $2.0<y<4.5$ using data collected by LHCb in the years 2016, 2017, and 2018. 
The data correspond to an integrated luminosity of $\mathcal{L}=4.18\pm0.08\invfb$. 
The \jpsi and \Upsilonres mesons are reconstructed in the dimuon final state.
The polarisation of the \jpsi and \Upsilonres mesons is assumed to be zero.
The polarisation of \JpsiUpsilon pairs is not yet measured, but all the LHC measurements so far indicate a small polarisation for quarkonia~\cite{ALICE:2011gej,CMS:2013gbz,CMS:2012bpf,LHCB-PAPER-2013-008,LHCB-PAPER-2013-067,LHCB-PAPER-2017-028}.

\section{Detector and dataset}
\label{sec:dataset}
The \lhcb detector~\cite{LHCb-DP-2008-001,LHCb-DP-2014-002} is a single-arm forward
spectrometer covering the \mbox{pseudorapidity} range $2<\eta <5$,
designed for the study of particles containing \bquark or \cquark
quarks. The detector includes a high-precision tracking system
consisting of a silicon-strip vertex detector surrounding the $pp$
interaction region, a large-area silicon-strip detector located
upstream of a dipole magnet with a bending power of about
$4{\mathrm{\,Tm}}$, and three stations of silicon-strip detectors and straw
drift tubes placed downstream of the magnet.
The tracking system provides a measurement of the momentum, \ptot, of charged particles with
a relative uncertainty that varies from 0.5\% at low momentum to 1.0\% at 200\gevc.
The minimum distance of a track to a primary $pp$ collision vertex (PV), the impact parameter, 
is measured with a resolution of $(15+29/\pt)\mum$,
where \pt is in\,\gevc.
Particle identification (PID) for charged hadrons uses information
from two ring-imaging Cherenkov detectors. 
Photons, electrons and hadrons are identified by a calorimeter system consisting of
scintillating-pad and preshower detectors, an electromagnetic
and a hadronic calorimeter. Muons are identified by
a system composed of alternating layers of iron and multiwire
proportional chambers.

The online event selection is performed by a trigger~\cite{LHCb-DP-2012-004}, 
which consists of a hardware stage, based on information from the calorimeter and muon
systems, followed by a software stage, which applies a full event
reconstruction.
For this analysis, data were collected with varying trigger conditions particularly for the hardware stage, which requires two muons with $\pt(\mu_1)\times\pt(\mu_2)>(1.3)^2,\,(1.5)^2 \text{ or } (1.8)^2\gevgevcc$. In the software trigger stage, at least two muons with $\pt >300\mevc$, $p>6\gevc$, and dimuon invariant mass $m(\mumu)>2.7\gevcc$ are required. 
Candidate \jpsi and \Upsilonres mesons are built from two oppositely charged tracks that are identified as muons and form a good vertex.
These muons from the \jpsi (\Upsilonres) decays are required to be in the kinematic region $2<\eta<5$, $ \pt>650 \text{ }(3000)\mevc$, and $p>6\gevc$. 
The decay vertices of the \jpsi and \Upsilonres mesons have to be consistent with the same PV. 
The invariant mass of the \jpsi candidates is required to be within~$[3.0,3.2]\gevcc$ and that of the \Upsilonres candidates within~$[9,11]\gevcc$.

Two independent simulated samples of \jpsi and \Upsilonres decays are used to model effects of the detector acceptance and the imposed selection requirements for \jpsi and \Upsilonres decays respectively. Primary $pp$ collisions are generated using Pythia~8~\cite{Sjostrand:2007gs} with a specific LHCb configuration~\cite{LHCb-PROC-2010-056}. Decays of unstable particles are simulated using
EvtGen~\cite{Lange:2001uf}, in which the final-state radiation is generated with PHOTOS~\cite{Golonka:2005pn}. 
The interaction of the generated particles with the detector, and its response,
are implemented using the \geant
toolkit~\cite{Allison:2006ve, *Agostinelli:2002hh} as described in Ref.~\cite{LHCb-PROC-2011-006}. 
All simulated events are reconstructed and selected in the same way as
the data.

\section{Cross-section determination}
\label{sec:xsec}
The \JpsiUpsilon production cross-section is measured according to
 \begin{equation}
 \label{equ:cross-section}
                \sigma(\JpsiUpsilon) = \frac{N_\mathrm{cor}}{\mathcal{L}\times\BF_{\jpsi \to \mup\mun}\times\BF_{\PUpsilon \to \mup\mun}},
            \end{equation}
where $N_{\mathrm{cor}}$ 
is the efficiency-corrected signal yield of the \JpsiUpsilon associated production, $\mathcal{B}_{\jpsi \to \mu^+ \mu^-}$ and  $\mathcal{B}_{\Upsilonres \to \mu^+ \mu^-}$ are the branching fractions of the $\jpsi \to \mumu$ and $\PUpsilon\to \mumu$ decays~\cite{PDG2022}, respectively, and $\mathcal{L}=4.18\pm0.08\invfb$ is the integrated luminosity determined with the van-der-Meer-scan method~\cite{vanderMeer:296752, LHCb-PAPER-2014-047}.

The total detection efficiency $\epsilon_\text{tot}$ for the offline selected \JpsiUpsilon candidates is factorised as
\begin{equation}
\label{equ:eff_tot}
                \epsilon_{\mathrm{tot}}=\epsilon_{\mathrm{acc}}\times  \epsilon_{\mathrm{rec\&sel}} \times  \epsilon_{\mathrm{PID}} \times  \epsilon_{\mathrm{trig}}\text{,}
            \end{equation}
where $\epsilon_{\mathrm{acc}}$ is the geometrical acceptance, $\epsilon_{\mathrm{rec\&sel}}$ is the reconstruction and selection (not including  PID requirements) efficiency of all the candidates within the geometrical acceptance, $\epsilon_{\mathrm{PID}}$ is the PID selection efficiency of candidates reconstructed and passing other selection requirements, 
and $\epsilon_{\mathrm{trig}}$ is the trigger selection efficiency of candidates passing all selection requirements. 
The first three terms on the right-hand side of Eq.~\ref{equ:eff_tot} are computed by multiplying the individual efficiencies for \jpsi and \Upsilonres candidates, \ie, 
\begin{equation}
    \epsilon_i=\epsilon_i^{\jpsi}\times\epsilon_i^{\PUpsilon},
\end{equation}
where \textit{i} denotes acc, rec\&sel, or PID.
For the trigger selection, it is required that either the \jpsi or \Upsilonres candidate meets the trigger requirements specified in Sec.~\ref{sec:dataset}; therefore the trigger efficiency $\epsilon_{\mathrm{trig}}$ can be factorised as
\begin{equation}
    \epsilon_\mathrm{trig}=1-(1-\epsilon_\mathrm{trig}^{\jpsi})(1-\epsilon_\mathrm{trig}^{\PUpsilon}).
\end{equation}
Each efficiency component is determined separately for \jpsi and \Upsilonres mesons as a function of their \pt and rapidity $y$.
The  acceptances, the reconstruction and selection efficiencies, and the trigger efficiencies are estimated using simulation, 
while the PID efficiencies are obtained using a data-driven method~\cite{LHCb-DP-2013-001}.

To extract the signal yield, an unbinned extended maximum-likelihood fit is performed to the two-dimensional \JpsiUpsilon invariant-mass distribution.
Defining the signal distribution to be $S$ and combinatorial-background distribution $B$,
the components in the model used to fit the data fall into three categories, each corresponding to the following cases: 
\begin{itemize}
    \item the signal $\JpsiUpsilon$ candidates ($S_\jpsi\times S_\Upsilonres$, 
    in which $S_\Upsilonres$ is $S_\OneS +  S_\TwoS + S_\ThreeS$),
    \item a true \jpsi (\Upsilonres) candidate associated with a combinatorial \Upsilonres (\jpsi) background candidate ($S_\jpsi\times B_\Upsilonres$ and $B_\jpsi\times S_\Upsilonres$),
    \item and a purely combinatorial background ($B_\jpsi\times B_\Upsilonres$).
\end{itemize}
 The two-dimensional probability density function for each component factorises into the product of two one-dimensional distributions for the \jpsi and \Upsilonres candidates. 
The signal invariant-mass distribution of the \jpsi meson and that of each \Upsilonres meson are described by a combination of a double-sided Crystal Ball (DSCB) function~\cite{Skwarnicki:1986xj} and a Gaussian function with a shared mean.
The DSCB power-law tail parameters, the ratio of the width of the Gaussian function to that of the Gaussian core in the DSCB function, and the relative fraction of the DSCB to the Gaussian function are fixed to values estimated from simulation. 
The ratios between the means of the \Upsilonres Gaussian functions are fixed to the ratios of the known \Upsilonres masses.
As the mass resolution of signals is linearly dependent on invariant mass, the width ratios of the Gaussian functions are also fixed to the known mass ratios~\cite{PDG2022}.
The mean value and the width of the Gaussian function for the $\OneS$ are free parameters in the fit.

\begin{figure}[!t]
          \centering
\includegraphics[width=.48\textwidth]{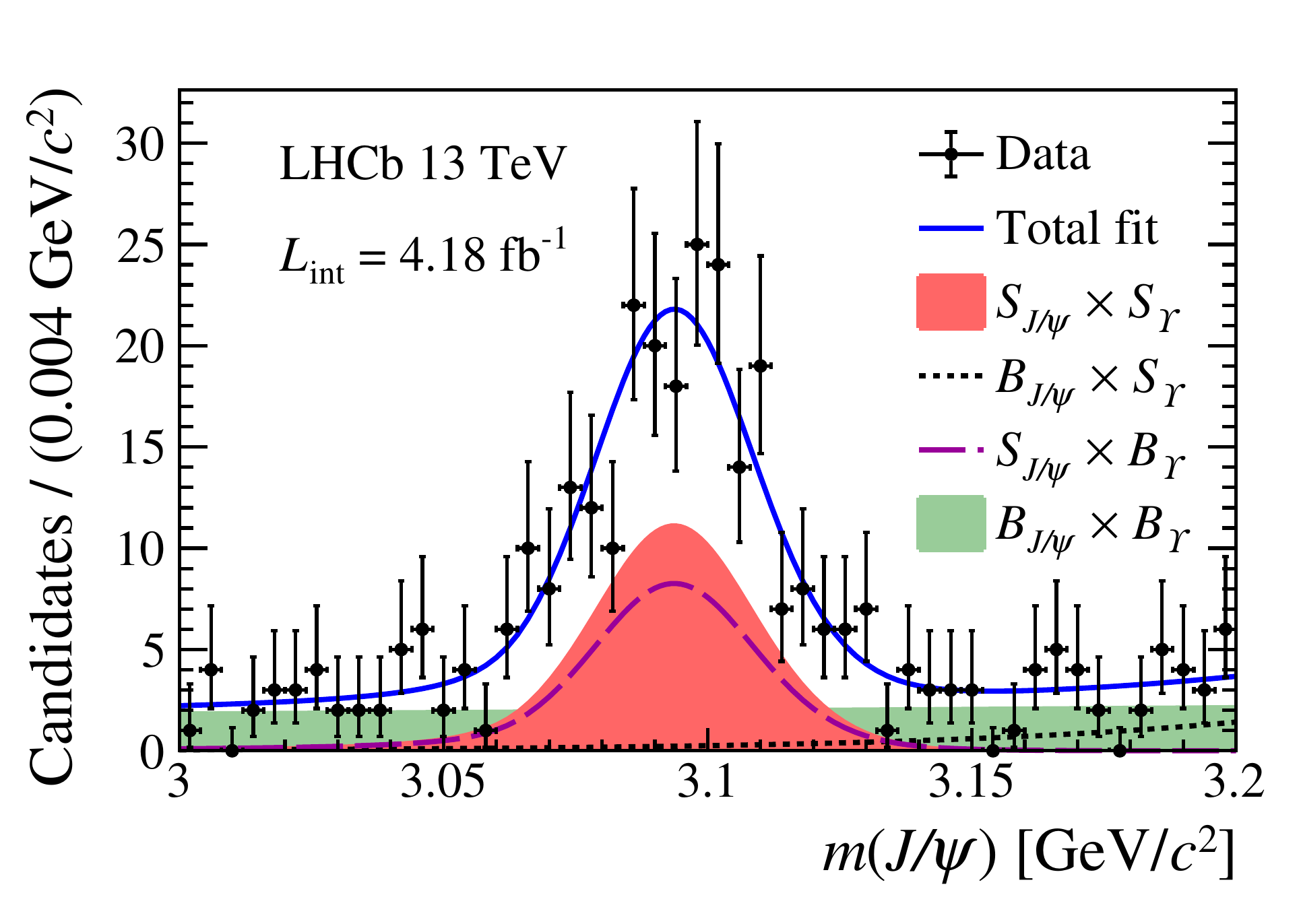}
\includegraphics[width=.48\textwidth]{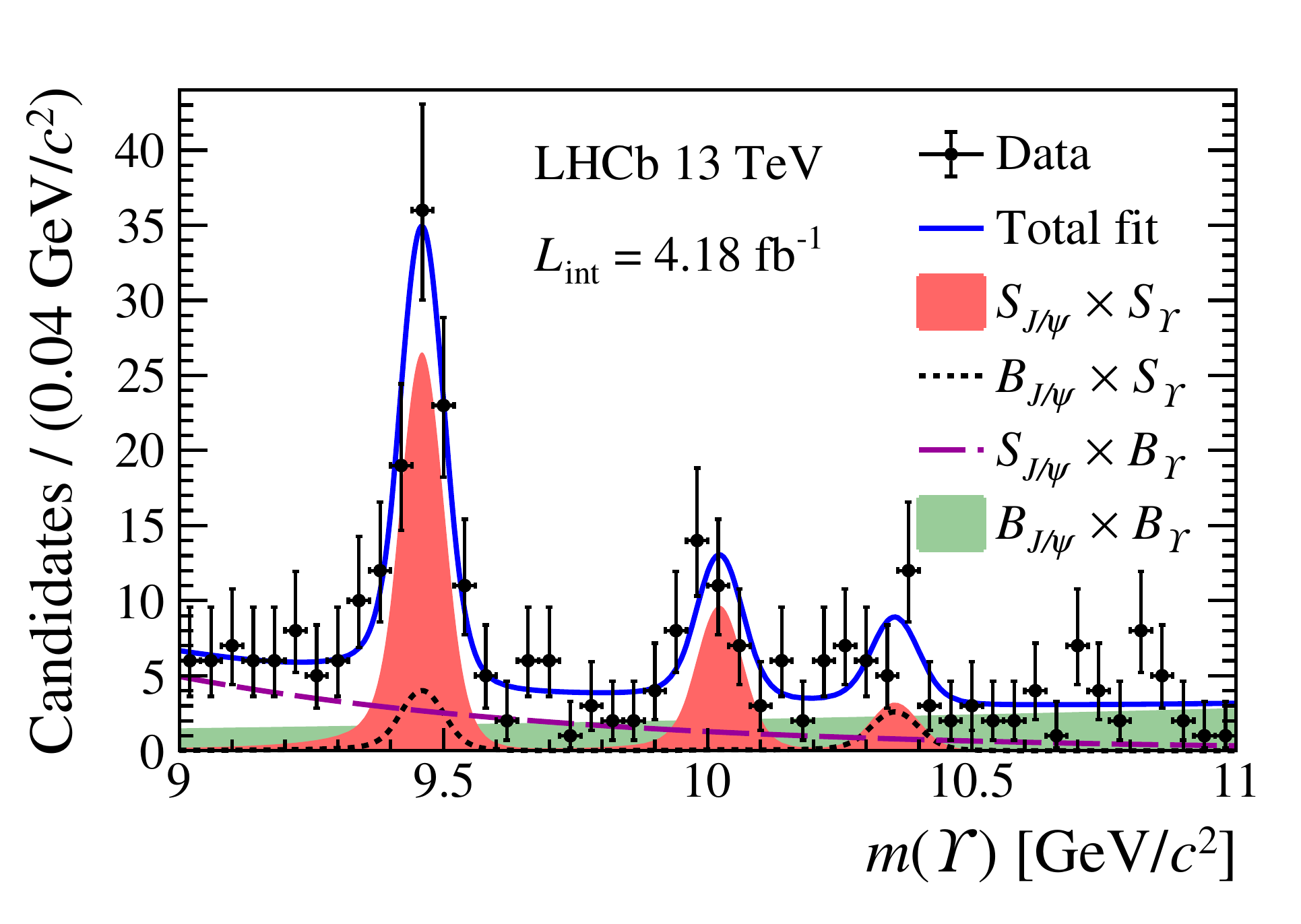}  
 \caption{Two-dimensional invariant mass distribution of the $\jpsi$--$\PUpsilon$ candidates projected on (left) the \jpsi invariant mass $m(\jpsi)$ and (right) the \Upsilonres invariant mass $m(\PUpsilon)$. 
 The black points with error bars represent the data. 
 The blue solid line represents the result of the fit to the data sample. The red shaded area represents the signal component. The black (dotted) and violet (dashed) lines represent the background where one dimuon candidate is a true meson decay and the other is combinatorial background. The green shaded area indicates the background where both the \jpsi and \Upsilonres candidates are combinatorial background.}
  \label{fig:data_fit}
 \end{figure}
Each combinatorial background is modelled by an exponential function.
The invariant-mass distributions of the  $\jpsi$ and $\Upsilonres$ candidates, together with the fit projections, are shown in Fig.~\ref{fig:data_fit}.
Associated $\jpsi$--$\OneS$ production is observed with a significance of \SigmaOne,
and evidence for $\jpsi$--$\TwoS$ production with a significance of \SigmaTwo is seen.
The total significance for the associated production of \jpsi with a combination of \OneS, \TwoS, and
\ThreeS~states is determined to be $8.5\,\sigma$. 
The significance is estimated by comparing the likelihood of the fit with and without the considered component in the fit, with relevant systematic uncertainties taken into consideration.
The raw yields of signals from the fit and their significances are summarised in Table~\ref{tab:yield}.
\begin{table}[t]
    \centering
        \caption{Raw yields from the fit, the efficiency-corrected yields~($N_\mathrm{cor}$) and the signal significances.}
    \begin{tabular}{cccc}
     Signal    &  Raw yields & $N_\mathrm{cor}$&Significances \\
     \hline
     \jpsi--$\OneS$    &$76\pm12$ &$\NcorOneSInclusive\pm\DeltaNcorOneSInclusive$ &\SigmaOne \\
     \jpsi--$\TwoS$    &$30\pm7$&$\NcorTwoSInclusive\pm\DeltaNcorTwoSInclusive$  &\SigmaTwo \\
     \jpsi--$\ThreeS$    &$10\pm6$ &-&$1.7\,\sigma$\\
    \end{tabular}
    \label{tab:yield}
\end{table}

Based on the fit result, a signal weight $w_i$ is calculated with the \sPlot method~\cite{Pivk:2004ty} for each $\JpsiUpsilon$ candidate, using the $\jpsi$ and $\Upsilonres$ invariant masses as  discriminating variables. 
These  weights are used to determine the efficiency corrected yield
$N^\text{cor}=\sum_j w_j/\epsilon_{\text{tot},\,j}$, where $j$ runs over all the $\JpsiUpsilon$ candidates in data. The total efficiency for the $j^{\text{th}}$ candidate, $\epsilon_{\text{tot},\,j}$, is defined in Eq.~\ref{equ:eff_tot} and determined as a function of the $\pt$ and $y$ of the \jpsi and \Upsilonres candidate.
The corrected yield, $N_{\text{cor}}$, is shown in Table~\ref{tab:yield}.

The signal extracted from the fit also contains  \jpsi mesons produced from  \bquark--hadron  decays (denoted \jpsi-from-\bquark).
The fraction of \jpsi-from-\bquark decays is determined to be 2.1\% 
using the prompt \jpsi and \jpsi-from-\bquark production cross-sections measured by LHCb in $\proton\proton$ collisions at $\sqrt{s}=13\tev$~\cite{LHCB-PAPER-2015-037}.
The difference between the average reconstruction and selection efficiencies of the two components is evaluated from simulation; the selection criteria constraining \jpsi and \Upsilonres decay vertices to the PV favour the prompt \jpsi candidates and suppress the \jpsi-from-\bquark candidates. 
With current measurements of $\sigma_{\mathrm{eff}}$, the $\JpsiUpsilon$ associated production is predicted to be dominated by the DPS process.
It is therefore assumed that in events with \jpsi and \Upsilonres candidates the relative cross-section of the \jpsi-from-\bquark production with respect to the prompt \jpsi production is the same as the relative cross-section of \jpsi-from-\bquark in inclusive \jpsi events. 
An uncertainty is assigned for systematic effects due to this assumption.
The \jpsi-from-\bquark component is subtracted when calculating the total cross-section of the associated $\jpsi$--$\PUpsilon$ production.

\section{Systematic uncertainties}
\label{sec:systematics}
Systematic uncertainties from various sources are studied and summarised in Table~\ref{tab:SU}. 
\begin{table}[ht!]
        \centering
        \caption{Relative systematic uncertainties on the measurement of the $\jpsi$--$\PUpsilon(1S)$ and $\jpsi$--$\TwoS$ cross-sections.}
        \begin{tabular}{lcc}
            Source                                       & $\sigma(\JpsiUpsilon(1S))$  (\%) &  $\sigma(\JpsiUpsilon(2S))$ (\%)\\
            \hline
            Signal shape                                    &2.0 &2.0 \\
            Limited simulation size               &0.7 &1.6 \\
            Tracking efficiency                             &3.3 &3.3\\
            PID efficiency                                       &0.5 &0.5\\
            \jpsi-from-\bquark fraction                                &1.1 &1.1\\
            Trigger efficiency                              &1.0 &1.0\\
            Radiative tail &2.0  &2.0\\
            Vertex fit &0.6 &0.6\\
            Luminosity                                      &1.9 &1.9\\
            $\mathcal{B}_{\jpsi \to \mu^{+}\mu^{-}}$        &0.6 &0.6\\
            $\mathcal{B}_{\OneS \to \mu^{+}\mu^{-}}$    &2.0 &-- \\
            $\mathcal{B}_{\TwoS \to \mu^{+}\mu^{-}}$    &-- &8.8 \\
            \hline
            Total                                           &5.5 &10.3\\
            \label{tab:SU}
        \end{tabular}
    \end{table}  
The total relative systematic uncertainty is 5.5\% for \SigOne and 10.3\% for \SigTwo production, corresponding to the quadratic sum of all the contributions discussed below.

The uncertainty introduced by the imperfect modelling of the signal shape is estimated by fitting the data with an alternative signal model.
The alternative shape is the shape of the simulated signal distribution, convolved with a Gaussian function to account for the different invariant-mass resolution observed between data and simulation. 
The 2.0\% relative change of the signal yield is assigned as the systematic uncertainty for both \SigOne and \SigTwo signals.

Due to the limited sizes of the simulation samples, the efficiencies obtained from simulation are subject to fluctuations. The statistical uncertainties of the efficiencies are propagated to the efficiency-corrected signal yield with a pseudoexperiment Monte Carlo method. 
The variations of the efficiency-corrected signal yields in pseudoexperiments are taken as the systematic uncertainty and are $0.7\%$ for $\jpsi$--$\OneS$ production and $1.6\%$ for $\jpsi$--$\TwoS$ production.

The track reconstruction efficiency as a function of muon $p$ and $\eta$ is estimated with $\jpsi\to\mumu$ decays in a calibration data sample using a tag-and-probe technique~\cite{LHCb-DP-2013-002}.  
The difference between data and simulation is used to calibrate the reconstruction efficiency estimated using simulated signal decays.
The uncertainties on these calibrations are propagated to the efficiency-corrected yield using a pseudoexperiment Monte Carlo method. The uncertainty is 0.9\% for both the \SigOne and \SigTwo signals. 
An additional uncertainty of 0.8\% per track is assigned to account for the difference in detector occupancy between data and simulation~\cite{LHCb-DP-2013-002}. A total relative uncertainty of 3.3\% is attributed to the tracking efficiency calibration.

The muon PID efficiency is calculated with $\jpsi\to\mumu$ decays using a data-driven tag-and-probe method~\cite{LHCb-DP-2013-001}.
The efficiency is estimated in bins of \pt and $\eta$ of the probe track, for which the binning scheme introduces an uncertainty. Alternative binning schemes are used to estimate variations of the PID efficiency. The relative change with respect to the default value,    0.5\%, is quoted as the systematic uncertainty.

The trigger efficiency estimated from simulation is compared with that from data with a data-driven method~\cite{LHCb-PUB-2014-039,LHCB-PAPER-2016-057}. 
A relative difference of 1.0\% is quoted as the systematic uncertainty.

The uncertainties introduced by the mismodelling of the radiative decay tails and the requirement on the vertex fit quality is estimated to be 2\% and 0.6\%, respectively, following Ref.~\cite{LHCB-PAPER-2015-037,LHCb-PAPER-2018-002} by comparing data and simulation. 

The fraction of \jpsi-from-\bquark is calculated from the \pt- and $y$-dependent \jpsi-from-\bquark fraction in single \jpsi production and the relative detection efficiency between \jpsi-from-\bquark and prompt production, convolved with the \jpsi kinematic distributions in simulation. 
The uncertainty of \jpsi kinematic distributions used to estimate the fraction and the possible difference of \jpsi-from-\bquark fraction between inclusive \jpsi production and associated \JpsiUpsilon production are sources of systematic uncertainties. 
The first uncertainty is evaluated using the maximum and minimum values of the inclusive \jpsi-from-\bquark fraction in the studied kinematic range of this analysis. The second source is estimated conservatively by varying the effective cross-section $\sigma_{\mathrm{eff}}$ of the  $\Upsilonres$--$B$ hadron associated production  by 50\%. In total, this systematic uncertainty is estimated to be 1.1\%. 
The effect of events where \jpsi and \Upsilonres mesons originate from different PVs is estimated to be negligible.

The relative uncertainty on the luminosity is estimated to be 1.9\%.
The relative uncertainties on the branching fractions $\mathcal{B}_{\jpsi \to \mu^+ \mu^-}$, $\mathcal{B}_{\OneS  \to \mu^+ \mu^-}$ and $\mathcal{B}_{\TwoS  \to \mu^+ \mu^-}$ are 0.6\%, 2.0\% and 8.8\%,  
respectively~\cite{PDG2022}. They are propagated to the cross-section uncertainties.

\section{Results and comparison to theory}
\label{sec:results}
The $\JpsiUpsilon$ production is measured in the fiducial region, $2<y<5$ and $\pt<10\gevc$ for the \jpsi meson,  $2<y<5$ and $\pt<30\gevc$ for the \Upsilonres meson. 
 The associated prompt-production cross-section of $\jpsi$--$\OneS$ is measured to be 
\begin{equation*}
                \sigma(\JpsiUpsilon(1S)) =\XsecOneS \pm \SigmaStatXsecOneS \pm \SigmaSystXsecOneS \pm \SigmaExtXsecOneS \pb\text{,}
            \end{equation*}
and the associated prompt-production cross-section of $\jpsi$--$\TwoS$ is measured to be                  \begin{equation*}               \sigma(\JpsiUpsilon(2S)) = \XsecTwoS \pm \SigmaStatXsecTwoS  \pm\SigmaSystXsecTwoS  \pm\SigmaExtXsecTwoS\pb\text{,}
            \end{equation*}
where the first uncertainties are statistical, the second systematic, and the third due to the uncertainties on the quarkonium decay branching fractions.

The ratio of the cross-sections of $\JpsiUpsilon(2S)$ and $\JpsiUpsilon(1S)$ multiplied by the respective branching ratio is calculated to be
     \begin{equation*}
          \frac{\mathcal{B}_{\PUpsilon(2S) \to \mu^{+}\mu^{-}}\times\sigma(\JpsiUpsilon(2S))}{\mathcal{B}_{\PUpsilon(1S) \to \mu^{+}\mu^{-}}\times\sigma(\JpsiUpsilon(1S))} = 0.442 \pm 0.143 \pm 0.004 \text{,}
     \end{equation*}
where the first uncertainty is statistical and the second systematic.
The result is consistent with the measurement
of  $\Upsilonres(1S,2S)$ production~\cite{LHCb-PAPER-2018-002} and of the  production of $\Upsilonres(1S, 2S)$ mesons associated with an open charm hadron
~\cite{LHCB-PAPER-2015-046}, as is expected by the DPS assumption.
     
\begin{figure}[tb]
      \centering 
    \begin{minipage}[c]{.38\textwidth} 
    \centering 
    \caption{Effective cross-sections measured in different particle production by various experiments.
    For the \lhcb measurements of $\JpsiUpsilon(1S)$ and $\JpsiUpsilon(2S)$ effective cross-sections, the error bars (grey boxes) represent the experimental (total) uncertainties.
    The result for $\jpsi$--$\jpsi$ production at LHCb is given as a range covering several model-dependent measurements~\cite{LHCB-PAPER-2016-057}.
    Results from other analyses are reproduced from Refs.~\cite{Chapon:2020heu,LHCB-PAPER-2015-046}.
    } 
    \label{fig:comparison} 
  \end{minipage}
  \begin{minipage}[c]{.6\textwidth} 
    \centering 
\includegraphics[width=0.9\textwidth]{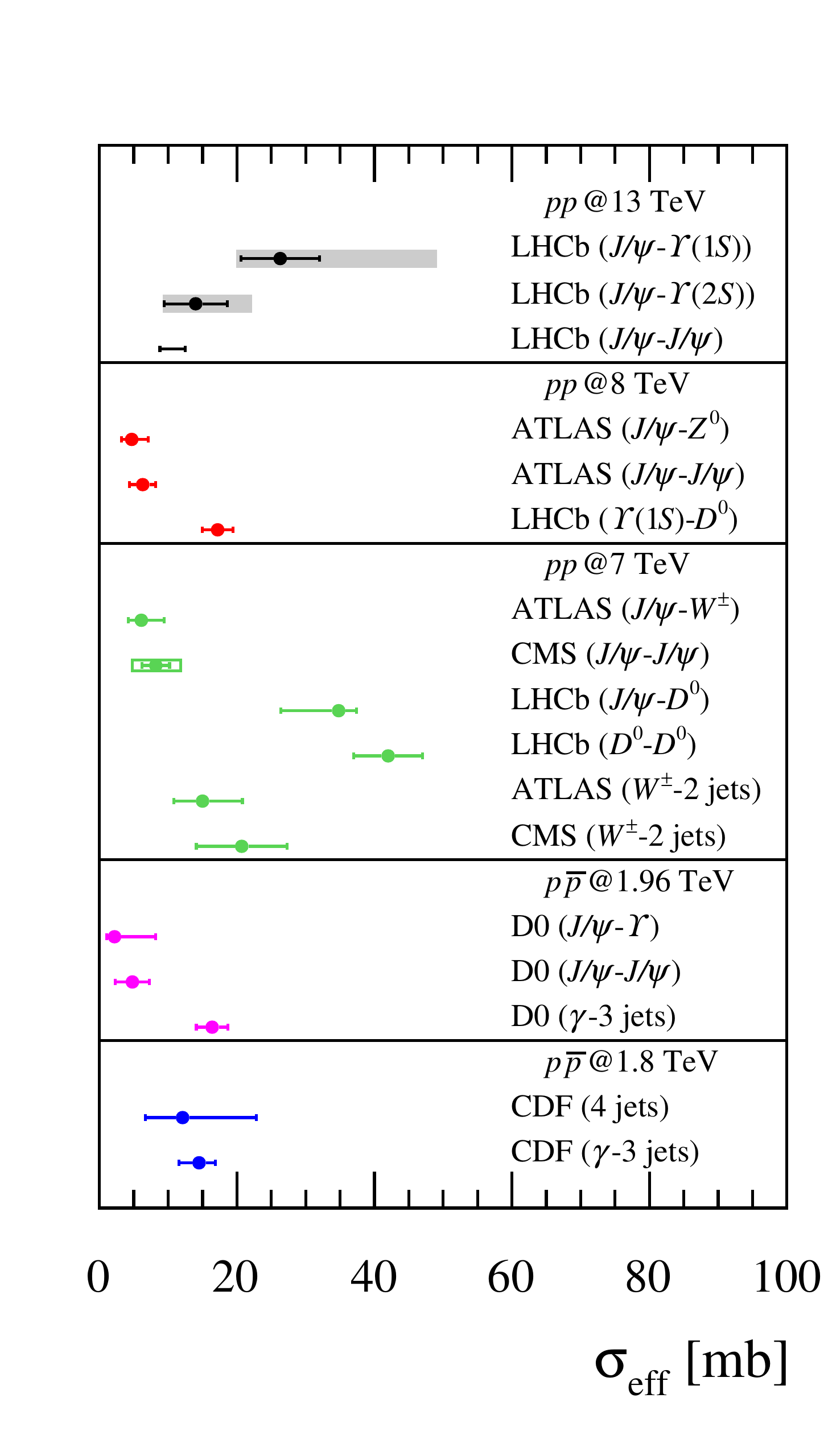}
  \end{minipage} 
\end{figure}

Using the DPS cross-section of the $\JpsiUpsilon$ associated production and the prompt-production cross-sections of single \jpsi and \Upsilonres mesons, the effective cross-section is calculated according to Eq.~\ref{equ:effective}.
The DPS cross-sections are estimated by subtracting the SPS cross-sections,  $20^{+52}_{-15}\pb$ for $\JpsiUpsilon(1S)$ and $8^{+22}_{-6}\pb$  for $\JpsiUpsilon(2S)$, from the associated-production cross-sections. The SPS cross-sections are adapted from the calculations with the NRQCD framework in Ref.~\cite{Shao:2016wor}.
The prompt-production cross-sections of \jpsi and \Upsilonres are taken from Ref.~\cite{LHCB-PAPER-2015-037} and Ref.~\cite{LHCb-PAPER-2018-002}, respectively. 
The effective cross-section is
        \begin{equation*}
                \sigma_{\mathrm{eff}}(\JpsiUpsilon(1S)) = 26 \pm 5 \pm 2 ^{+22}_{-3}\mbarn,
            \end{equation*}
            \begin{equation*}
                \sigma_{\mathrm{eff}}(\JpsiUpsilon(2S)) = 14 \pm 5 \pm 1 ^{+7}_{-1} \mbarn,
            \end{equation*}
where the first uncertainties are statistical, the second uncertainties are systematic, and the last terms are uncertainties due to theory calculations.
The results of  $\sigma_{\mathrm{eff}}$ 
are broadly compatible with measurements using hadroproduction of other particles, as shown in Fig.~\ref{fig:comparison}. The effective cross-sections for $\JpsiUpsilon(1S)$ and $\JpsiUpsilon(2S)$ production are consistent. 

\begin{figure}[tb]
    \centering
    \includegraphics[width = 0.45\textwidth]{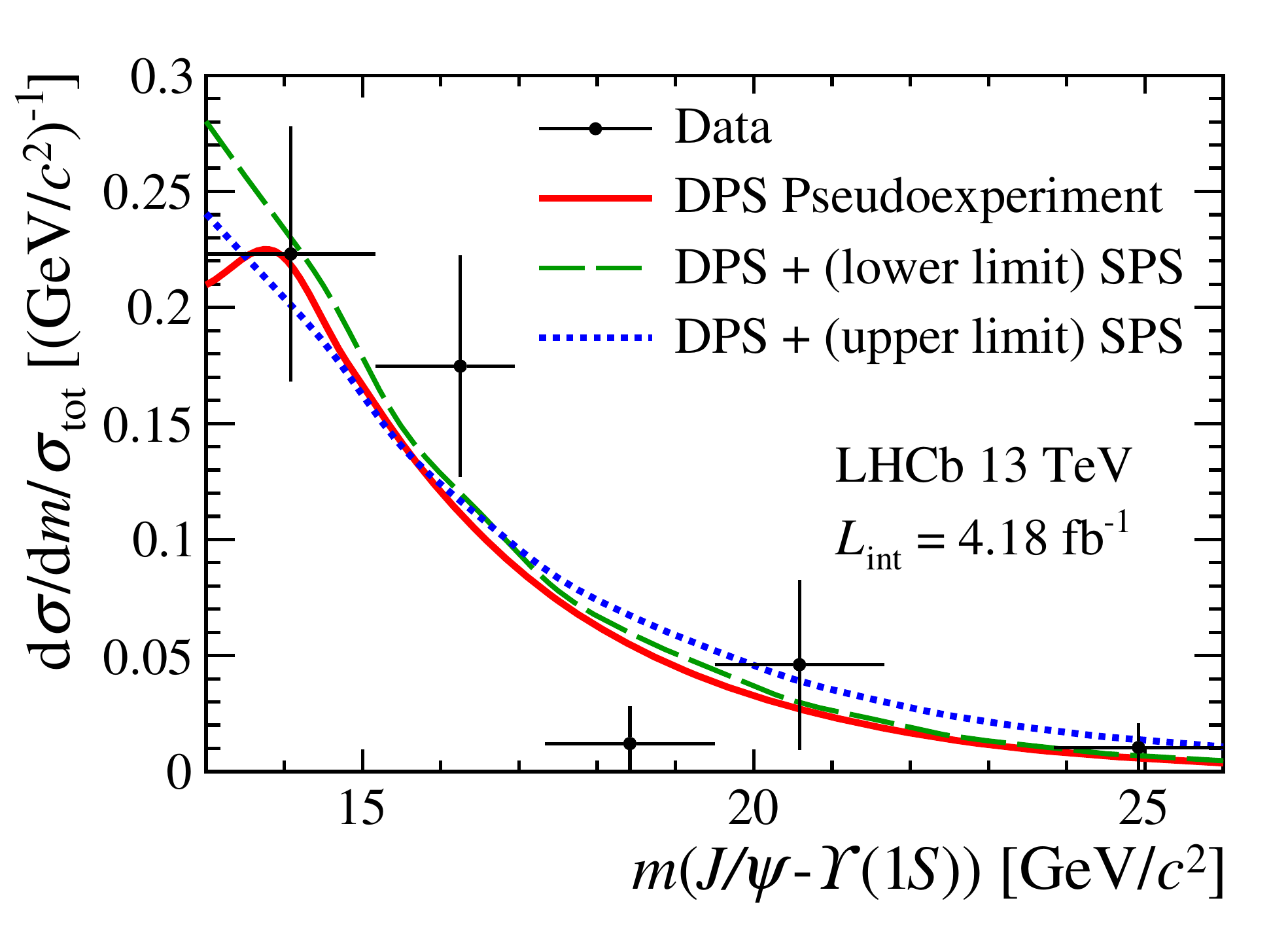}
    \includegraphics[width = 0.45\textwidth]{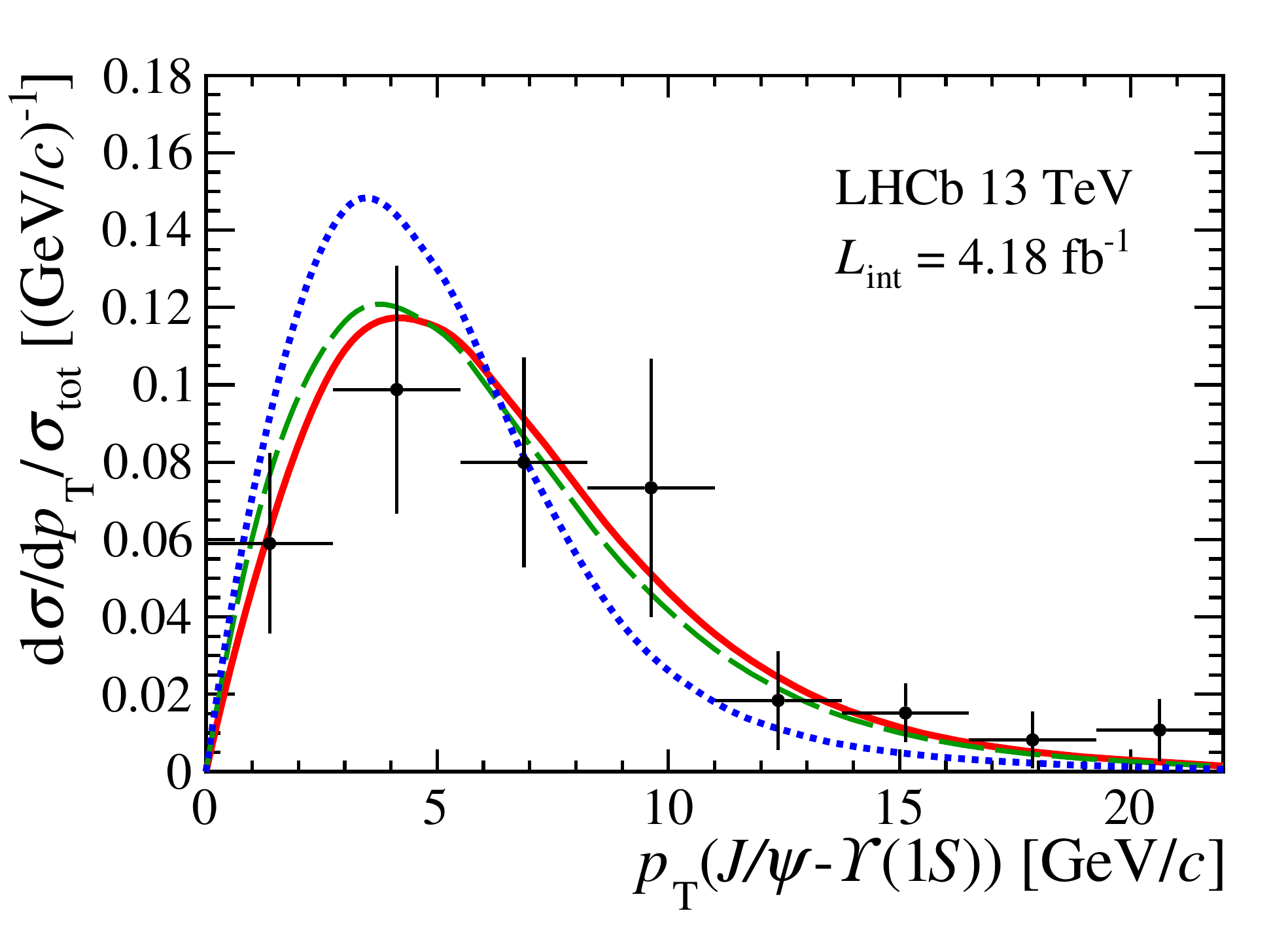}\\
    \includegraphics[width = 0.45\textwidth]{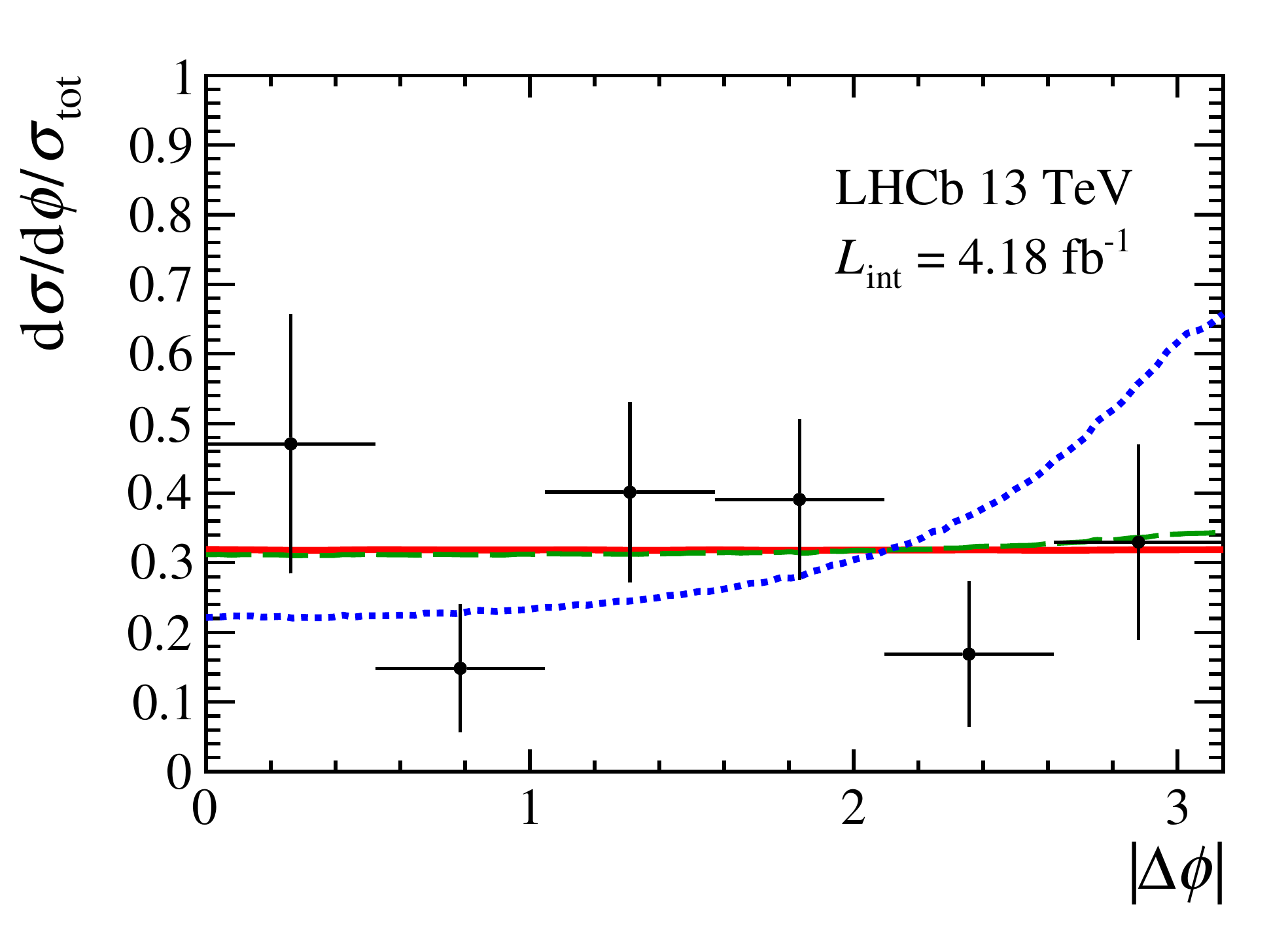}
    \includegraphics[width = 0.45\textwidth]{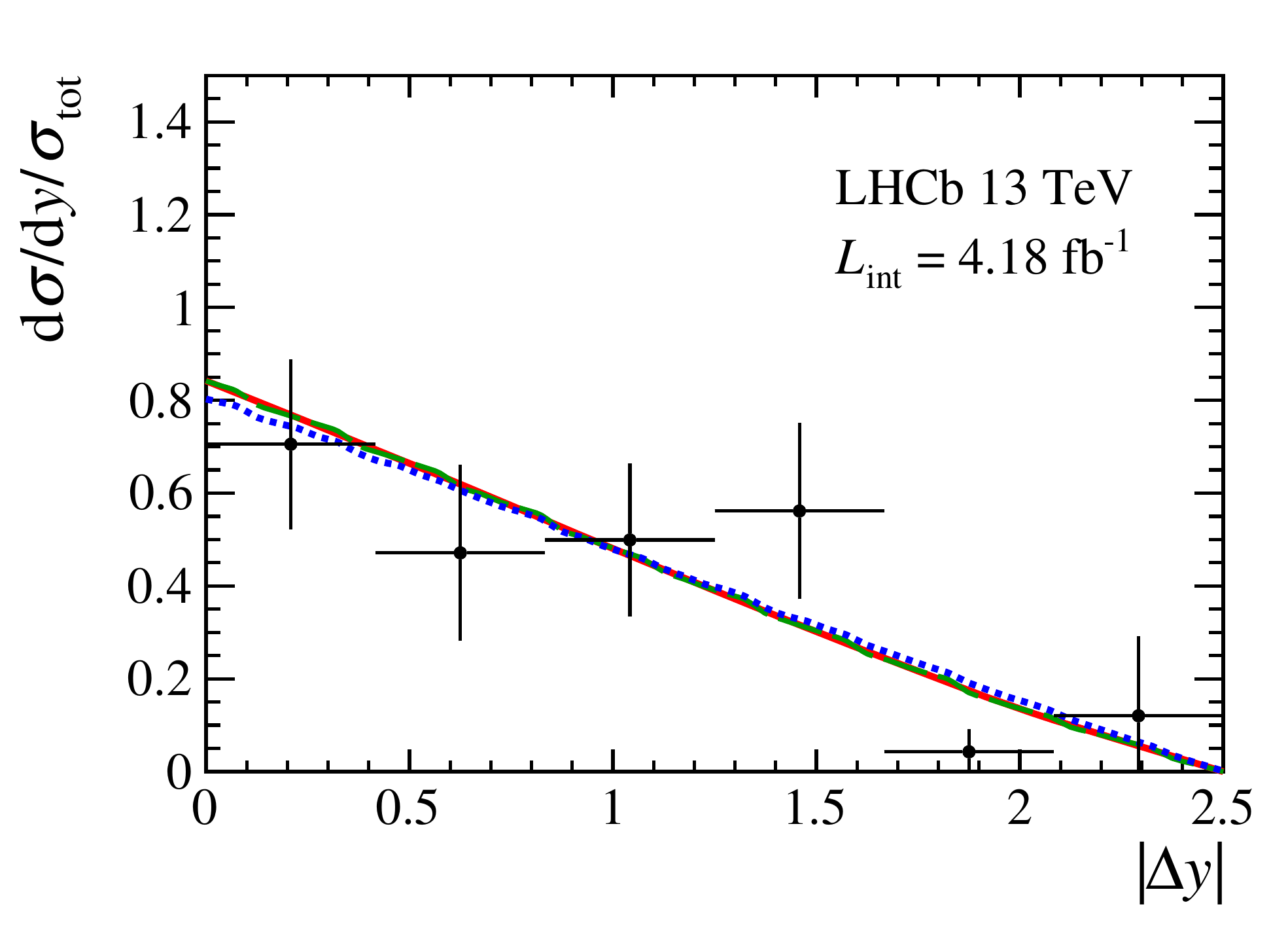}\\
    \includegraphics[width = 0.45\textwidth]{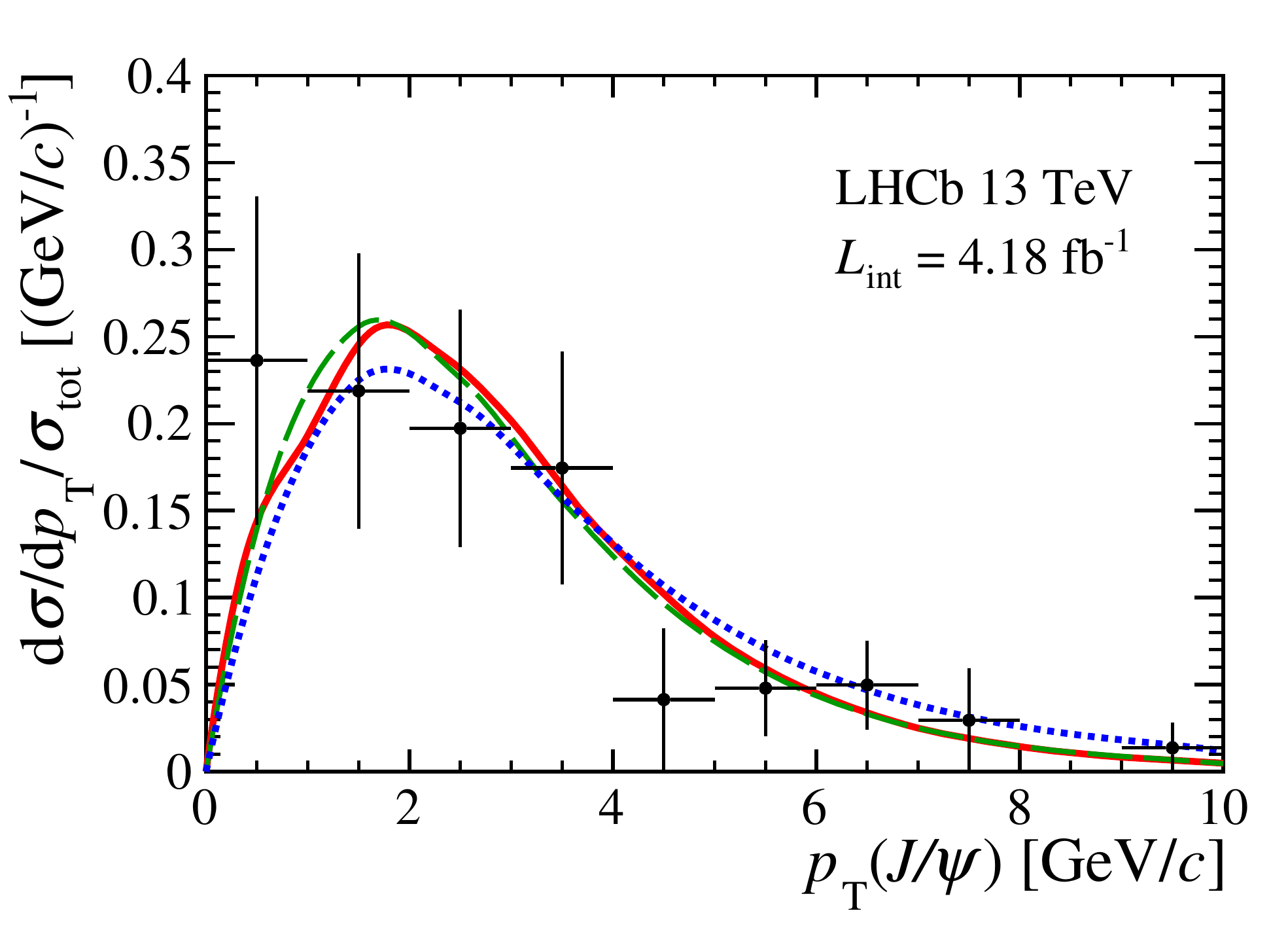}
    \includegraphics[width = 0.45\textwidth]{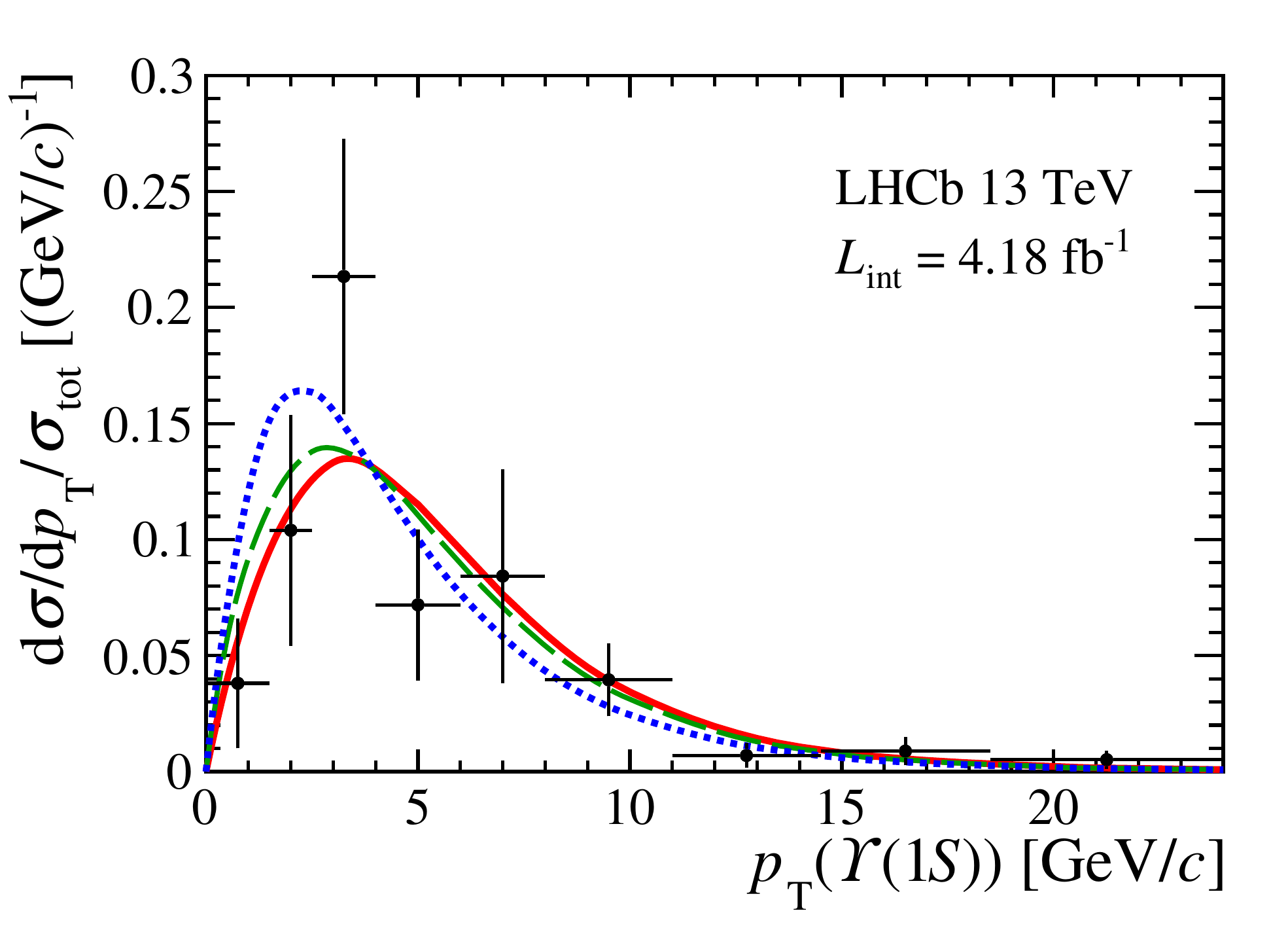}\\
    \caption{Distributions of (top left) invariant mass  $m(\JpsiUpsilon(1S))$,  (top right) transverse momentum $\pt(\JpsiUpsilon(1S))$, (middle left)  azimuthal angle difference $|\Delta \phi|$, (middle right) rapidity difference $|\Delta y|$ of $\jpsi$--$\OneS$, and (bottom) $\pt$ of the \jpsi and \OneS mesons for data and the pseudodata modelled from previous measurements~\cite{LHCB-PAPER-2015-037,LHCb-PAPER-2018-002}.  Black points with error bars represent the data distributions with statistical uncertainties. Red solid lines represent distributions of the DPS pseudodata. Green dashed lines and blue dotted lines represent the DPS distribution together with the lower limit and the upper limit of the SPS distributions, respectively.}
    \label{fig:distributions}
\end{figure}

The distributions of some observables related to both $\jpsi$ and $\OneS$ mesons are studied to probe the kinematic correlation between the two mesons. These variables include the $\jpsi$-$\OneS$ invariant mass, the  transverse momentum of the $\jpsi$--$\OneS$ system, the relative azimuthal angle, and the rapidity difference between \jpsi and $\OneS$ mesons.
The distributions of these variables and the transverse momentum distributions of the $\jpsi$ and $\OneS$ mesons are presented in Fig.~\ref{fig:distributions}, after correcting for the total detection efficiency  and subtracting  the background contribution.
The large uncertainties in the correlation distributions are due to the small size of the signal data sample, which dominate over systematic uncertainties. These distributions are compared with pseudoexperiments corresponding to different contributions of the DPS and SPS processes.
Since the \jpsi and \Upsilonres mesons are produced independently from two different parton-parton scatterings in the DPS processes, the kinematic distribution of each meson in the associated production  is expected to be similar to that in the single meson production.
As a result, using the measured $\pt$ and $y$ distributions in single-\jpsi and -\OneS production~\cite{LHCB-PAPER-2015-037,LHCb-PAPER-2018-002}, the DPS distributions of \jpsi and \OneS kinematic variables in the associated production can be modelled. 
The distributions for the SPS production are taken from calculations in Ref.~\cite{Shao:2016wor}. Three samples of pseudodata are generated, differing in the relative SPS to DPS fractions. For one sample, only DPS production is considered. For the other two samples, the SPS fraction is calculated using either the upper or lower limit of the predicted cross-section in Ref.~\cite{Shao:2016wor}, and the DPS cross-section is the \lhcb measurement of the total cross-section with the SPS cross-section subtracted.
As shown in Fig.~\ref{fig:distributions}, the \jpsi and \OneS $\pt$ distributions and the kinematic correlations between the \jpsi and \OneS mesons in the  three pseudodata samples are all found to be consistent with measurements.

\section{Summary}
\label{sec:summary}

Using $\proton\proton$ collision data collected by the LHCb experiment at a centre-of-mass energy of $13\tev$ corresponding to an integrated luminosity of $4.18\pm0.08\invfb$, a signal for \SigOne production is observed with a significance of \SigmaOne, and evidence is found for \SigTwo production.
Within the fiducial region $2.0<y<4.5$ and $\pt<10\text{ } (30)\gevc$ for \jpsi (\Upsilonres) mesons, the cross-section for \SigOne production is measured to be \hbox{$\XsecOneS \pm \SigmaStatXsecOneS \stat\pm \SigmaSystXsecOneS \syst \pm \SigmaExtXsecOneS \,(\BF) \pb$} and that for \SigTwo production  $\XsecTwoS \pm \SigmaStatXsecTwoS \stat \pm\SigmaSystXsecTwoS \syst \pm \SigmaExtXsecTwoS \,(\BF) \pb$. 
The effective cross-sections for \SigOne and \SigTwo production are calculated and found to be compatible with previous measurements in various processes. 
Kinematic correlations between the \jpsi and $\OneS$ mesons are studied, showing results consistent with both the hypotheses that the associated production is dominated by the DPS process or has  DPS and SPS contributions.
In the future, further data from LHCb combined with improved theory calculations will help constrain the associated quarkonium production mechanism.

% Do not include this in any draft (just for information in the template)
%\input{acknowledgements_intro}
% Comment this in for paper drafts; do not include this in analysis note, conference and figure reports
\section*{Acknowledgements}
%
% These Acknowledgements valid from 3-May-2019
%
\noindent We express our gratitude to our colleagues in the CERN
accelerator departments for the excellent performance of the LHC. We
thank the technical and administrative staff at the LHCb
institutes.
We acknowledge support from CERN and from the national agencies:
CAPES, CNPq, FAPERJ and FINEP (Brazil); 
MOST and NSFC (China); 
CNRS/IN2P3 (France); 
BMBF, DFG and MPG (Germany); 
INFN (Italy); 
NWO (Netherlands); 
MNiSW and NCN (Poland); 
MEN/IFA (Romania); 
%MSHE (Russia); 
MICINN (Spain); 
SNSF and SER (Switzerland); 
NASU (Ukraine); 
STFC (United Kingdom); 
DOE NP and NSF (USA).
We acknowledge the computing resources that are provided by CERN, IN2P3
(France), KIT and DESY (Germany), INFN (Italy), SURF (Netherlands),
PIC (Spain), GridPP (United Kingdom), 
%RRCKI and Yandex LLC (Russia), 
CSCS (Switzerland), IFIN-HH (Romania), CBPF (Brazil),
Polish WLCG  (Poland) and NERSC (USA).
We are indebted to the communities behind the multiple open-source
software packages on which we depend.
Individual groups or members have received support from
ARC and ARDC (Australia);
Minciencias (Colombia);
AvH Foundation (Germany);
EPLANET, Marie Sk\l{}odowska-Curie Actions and ERC (European Union);
A*MIDEX, ANR, IPhU and Labex P2IO, and R\'{e}gion Auvergne-Rh\^{o}ne-Alpes (France);
Key Research Program of Frontier Sciences of CAS, CAS PIFI, CAS CCEPP, 
Fundamental Research Funds for the Central Universities, 
and Sci. \& Tech. Program of Guangzhou (China);
%Key Research Program of Frontier Sciences of CAS, CAS PIFI,
%Thousand Talents Program, and Sci. \& Tech. Program of Guangzhou (China);
%RFBR, RSF and Yandex LLC (Russia);
GVA, XuntaGal, GENCAT and Prog.~Atracci\'on Talento, CM (Spain);
SRC (Sweden);
the Leverhulme Trust, the Royal Society
 and UKRI (United Kingdom).

%\input{supplementary}

%\input{appendix}

% This should be taken out in the final paper
%\input{supplementary-app}

\addcontentsline{toc}{section}{References}
%\setboolean{inbibliography}{true}
\bibliographystyle{LHCb}
\bibliography{main,standard,LHCb-PAPER,LHCb-CONF,LHCb-DP,LHCb-TDR}

\newpage

% LHCb collaboration author list
% Data extracted on April 27th, 2023 at 12:37pm for paper reference LHCb-PAPER-2022-047
\centerline
{\large\bf LHCb collaboration}
\begin
{flushleft}
\small
R.~Aaij$^{32}$\lhcborcid{0000-0003-0533-1952},
A.S.W.~Abdelmotteleb$^{50}$\lhcborcid{0000-0001-7905-0542},
C.~Abellan~Beteta$^{44}$,
F.~Abudin{\'e}n$^{50}$\lhcborcid{0000-0002-6737-3528},
T.~Ackernley$^{54}$\lhcborcid{0000-0002-5951-3498},
B.~Adeva$^{40}$\lhcborcid{0000-0001-9756-3712},
M.~Adinolfi$^{48}$\lhcborcid{0000-0002-1326-1264},
P.~Adlarson$^{77}$\lhcborcid{0000-0001-6280-3851},
H.~Afsharnia$^{9}$,
C.~Agapopoulou$^{13}$\lhcborcid{0000-0002-2368-0147},
C.A.~Aidala$^{78}$\lhcborcid{0000-0001-9540-4988}
% ...

{\footnotesize \it

% ...
$^{9}$Universit{\'e} Clermont Auvergne, CNRS/IN2P3, LPC, Clermont-Ferrand, France\\
% ...
$^{13}$LPNHE, Sorbonne Universit{\'e}, Paris Diderot Sorbonne Paris Cit{\'e}, CNRS/IN2P3, Paris, France\\
% ...
$^{32}$Nikhef National Institute for Subatomic Physics, Amsterdam, Netherlands\\
% ...
$^{40}$Instituto Galego de F{\'\i}sica de Altas Enerx{\'\i}as (IGFAE), Universidade de Santiago de Compostela, Santiago de Compostela, Spain\\
% ...
$^{44}$Physik-Institut, Universit{\"a}t Z{\"u}rich, Z{\"u}rich, Switzerland\\
% ...
$^{48}$H.H. Wills Physics Laboratory, University of Bristol, Bristol, United Kingdom\\
% ...
$^{50}$Department of Physics, University of Warwick, Coventry, United Kingdom\\
% ...
$^{53}$School of Physics and Astronomy, University of Glasgow, Glasgow, United Kingdom\\
$^{54}$Oliver Lodge Laboratory, University of Liverpool, Liverpool, United Kingdom\\
% ...
$^{62}$Syracuse University, Syracuse, NY, United States\\
% ...
$^{77}$Department of Physics and Astronomy, Uppsala University, Uppsala, Sweden, associated to $^{53}$\\
$^{78}$University of Michigan, Ann Arbor, MI, United States, associated to $^{62}$\\
\bigskip
% ...
}
\end{flushleft}

\end{document}